\documentclass[twocolumn,floatfix,prb,aps,showpacs,superscriptaddress]{revtex4-1} 
\bibpunct{}{}{,}{n}{}{}

\usepackage{graphicx,epsf,amsmath,amssymb}
\usepackage{dcolumn}
\usepackage{bm, bbm}
\usepackage{enumerate}
\usepackage[usenames,dvipsnames,svgnames,table]{xcolor}
\usepackage{verbatim}

\usepackage{hyperref}
\hypersetup{
    bookmarks=false,         
    unicode=false,          
    pdftoolbar=true,        
    pdfmenubar=true,        
    pdffitwindow=false,     
    pdfstartview={FitH},    
    pdftitle={impurities in 3D quadratic band-touching Luttinger semimetals. Friedel and RKKY oscillations.},    
    pdfauthor={UMontreal},     
    pdfsubject={Report},   
    pdfcreator={UMontreal},   
    pdfproducer={UMontreal}, 
    pdfkeywords= {luttinger} {friedel} {rkky} {magnetic} {susceptibility} 
    pdfnewwindow=true,      
    colorlinks=true,       
    linkcolor=blue,          
    citecolor=red,        
    filecolor=magenta,      
    urlcolor=cyan           
}

\begin{document}

\preprint{APS/123-QED}

\title{Impurities in 3D quadratic band-touching Luttinger semimetals :\\ Friedel and RKKY oscillations}

\author{L. J. Godbout}
\affiliation{D\'epartement de Physique, Universit\'e de Montr\'eal, Montr\'eal, Qu\'ebec, H3C 3J7, Canada}

\author{S. Tchoumakov}
\affiliation{D\'epartement de Physique, Universit\'e de Montr\'eal, Montr\'eal, Qu\'ebec, H3C 3J7, Canada}

\author{W.~Witczak-Krempa}
\affiliation{D\'epartement de Physique, Universit\'e de Montr\'eal, Montr\'eal, Qu\'ebec, H3C 3J7, Canada}
\affiliation{Centre de Recherches Math\'ematiques, Universit\'e de Montr\'eal; P.O. Box 6128, Centre-ville Station; Montr\'eal (Qu\'ebec), H3C 3J7, Canada}
\affiliation{Regroupement Qu\'eb\'ecois sur les Mat\'eriaux de Pointe (RQMP)}

\date{\today}

\begin{abstract}
We investigate the response of 3D Luttinger semimetals to localized charge and spin impurities as a function of doping. The strong spin-orbit coupling of these materials strongly influences the Friedel oscillations and RKKY interactions. This can be seen at short distances with an $1/r^4$ divergence of the responses, and  anisotropic behavior.  
Certain of the spin-orbital signatures are robust to temperature, even if the charge and spin oscillations are smeared out, and give an unusual diamagnetic Pauli susceptibility. We compare our results to the experimental literature on the bismuth-based half-Heuslers such as YPtBi and on the pyrochlore iridate Pr$_2$Ir$_2$O$_7$.
\end{abstract}


\maketitle

\section{Introduction}

Luttinger semimetals [\cite{Luttinger}] such as HgTe, $\alpha-$Sn, YPtBi [\cite{chenglong}] and Pr$_2$Ir$_2$O$_7$ [\cite{kondo}] play an important role in the field of 3D topological materials. Their conduction and valence bands meet quadratically at a time-reversal invariant and inversion-symmetric point. This degeneracy can be lifted, for example by applying strain, to obtain a Dirac semimetal [\cite{taekoo,luttodirac}] or a topological insulator [\cite{luttoti}]. In these materials, the non-trivial topology is responsible for unusual magnetic [\cite{cmefinitesize, theoopticalcme, chiralanomalyoptics}], surface [\cite{thfermiarcs, fermiarcexp1, fermiarcexp2}] and transport [\cite{cmefinitesize, thmagnetores, weylmagnetores}] properties, that can also be met in Luttinger semimetals [\cite{spincur}]. In particular, a way to probe the strong spin-orbit properties of these materials is through their response to charge and spin impurities, respectively reffered to as Friedel oscillations [\cite{friedel,giuliani}] and RKKY interactions [\cite{rudermankittel, kasuya, yosida}].

The Friedel oscillations and RKKY interactions are a consequence of the sharp Fermi surface of the conduction electrons. In a 3D normal electron gas they typically scale as $\cos(2k_F r)/r^3$ where $r$ is the distance from the impurity and $k_F$ the Fermi wavevector [\cite{white}]. But in general, this response depends on dimension [\cite{giuliani,kurilovich2016,kurilovich2017}], band dispersion [\cite{rkkydirac1,rkkydirac2,rkkydirac3,rkkydirac4,bilayer,friedeldirac,friedelnodalline}] and temperature [\cite{rkkytemp1,rkkytemp2}]. For example, in Dirac and Weyl semimetals these responses decay faster at large separations, $r^{-5}$, [\cite{rkkydirac2,rkkydirac4,hosseini}] when the carrier density vanishes.
The quadratic dispersion of Luttinger semimetals leads to a slower decay, which may prove useful to explore the consequences of the strong spin-orbit interaction. Also, on the contrary to a Luttinger metal with heavy and light carriers [\cite{verma2019}], the chemical potential can serve to explore the spin-orbit coupling of each band separately.

In this work we study the Friedel oscillations and RKKY coupling of three-dimensional Luttinger semimetals at finite doping. This is inspired by recent experimental results on the bismuth based half-Heuslers [\cite{chenglong}] and the tentalizing phase diagram of the pyrochlore iridate Pr$_2$Ir$_2$O$_7$ [\cite{kondo, balicas, nakatsuji}]. We compute the charge and spin response at zero temperature analytically, and numerically at finite temperature. Also, because it was recently shown that Luttinger semimetals have a paramagnetic Landau susceptibility [\cite{leahy}], which is opposite to what is expected for a normal electron gas, we compute the Pauli susceptibility and find that it is diamagnetic.

This work is organized as follows. In Sec.~\ref{sec:model} we introduce and discuss the underlying model of a Luttinger semimetal. Sec.~\ref{sec:inhom} contains our main results, with the expression of the charge and of the spin response to a localized inhomogeneity. We discuss our results at zero temperature and as a function of temperature, and compute the spin susceptibility of a Luttinger semimetal. In Sec.~\ref{sec:discussion}, we compare our results to the existing litterature on Luttinger semimetals like bismuth-based half-Heuslers (YPtBi, LuPtBi, ...) and the pyrochlore Pr$_2$Ir$_2$O$_7$. We compare our results with the known literature for Dirac semimetals. 

\section{Model}
\label{sec:model}
\begin{figure}[tb]\textbf{}
    \centering
    \includegraphics[width = \columnwidth]{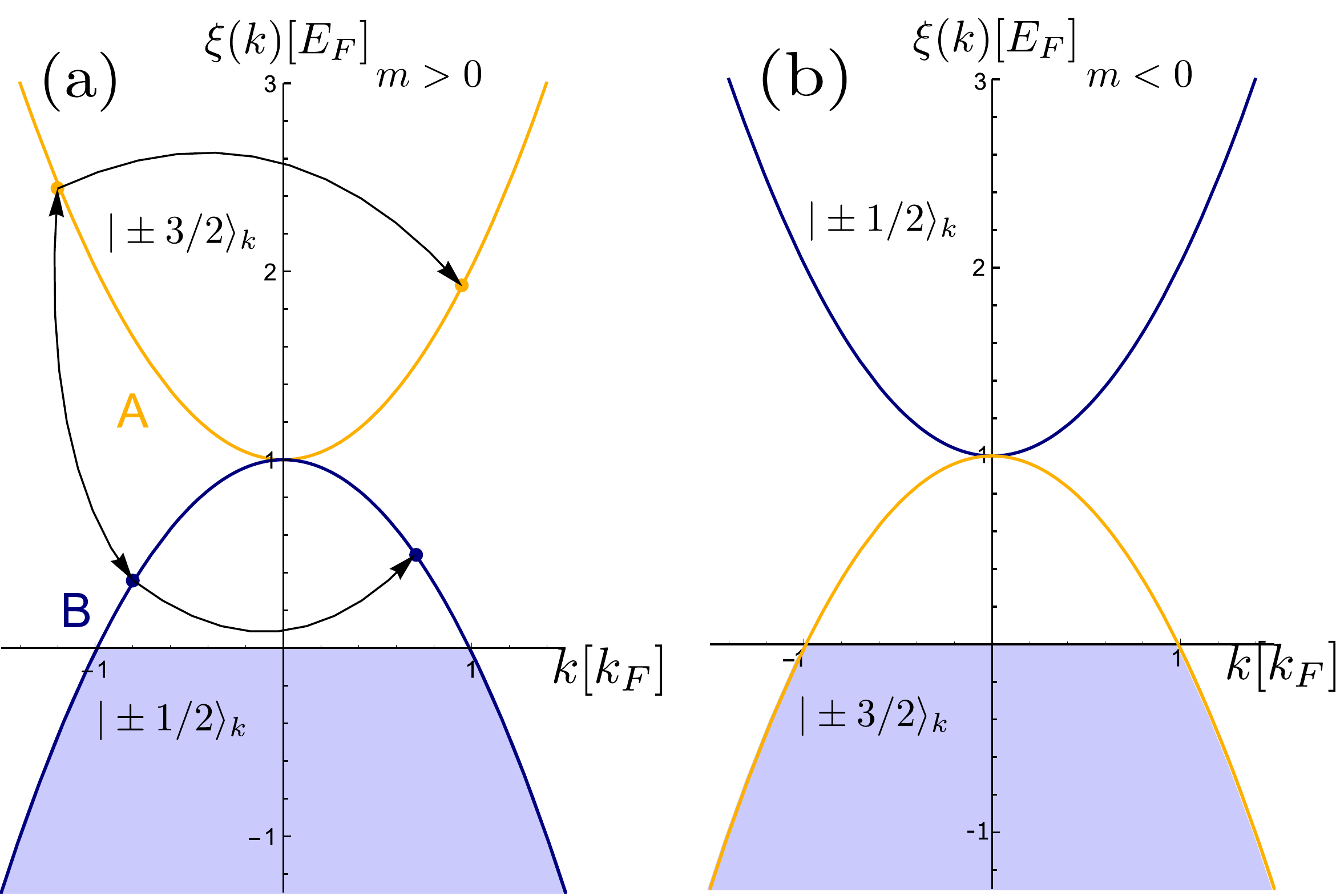}
    \caption{Spectrum of a Luttinger semimetal 
    for (a) $m > 0$ and (b) $m < 0$. The Fermi surface is in the lower band and, depending on the sign of the band mass $m$, it crosses different eigenstates (a) $|\pm 1/2\rangle_{\bf k}$ and (b) $|\pm 3/2\rangle_{\bf k}$ of the helicity operator $\hat{\lambda} = {\bf k}\cdot\hat{\bf J}/k$. This influences the intraband and interband scattering in the presence of a magnetic impurity. We respectively refer to the lower and upper band as the conduction and valence band in the text, since we consider hole doping.}
    \label{fig:spctr}
\end{figure}

At a quadratic band touching, the behavior of non-interacting electrons can be described with the Luttinger Hamiltonian [\cite{Luttinger}]
\begin{align}\label{eq:h0}
    \hat{H}_0({\bf k}) = \frac{\hbar^2}{2m} \left[  - \frac54 {\bf k}^2 \hat{\mathbbm{1}} + \left({\bf k}\cdot \hat{{\bf J}}\right)^2 \right]  - E_F,
\end{align} 
where the band mass is $m$ and $\hat{{\bf J}} = (\hat{J}_x,\hat{J}_y,\hat{J}_z)$ are the $j = 3/2$ total angular momentum operators. This model has rotation, inversion and time-reversal symmetries~[\cite{Luttinger}]. The 4 eigenstates of $\hat{H}_0({\bf k})$ can be labelled with by the eigenvalues $\lambda = \pm1/2, \pm3/2$ of the helicity operator $\hat{\lambda} = {\bf k}\cdot\hat{\bf J}/k$ and the corresponding spectrum $\xi_{\pm}({\bf k}) = \pm \hbar^2 k^2/(2m) - E_F$ is drawn in Fig.~\ref{fig:spctr}. The precise expression of the corresponding eigenvector is cumbersome and in the following we use the thermal Green's function
\begin{align}\label{eq:green}
    \hat{G}({\bf k}, i\omega_n) = -\frac{i\omega_n + \hat{H}_0({\bf k})}{(\xi_{+}({\bf k}) + i\omega_n)(\xi_{-}({\bf k}) + i\omega_n)},
\end{align}
where $\hbar \omega_n = (2n+1)\pi k_B T$ are the Matsubara frequency at temperature $T$.

In the following we set $\hbar = k_B = 1$, write energies in units of the Fermi energy, $|E_F|$, and wavevectors in units of the Fermi wavevector, $k_F$. We consider that the Fermi energy is in the lower band, $E_F = -k_F^2/(2|m|) < 0$, but we should have in mind that, depending on the sign of $m$, the Fermi surface can cross eigenstates with either helicity $\pm 1/2$ if $m > 0$ or $\pm 3/2$ if $m < 0$ (see Fig.~\ref{fig:spctr}). These two situations will alter the RKKY coupling.

\section{Response to charge and spin impurities}
\label{sec:inhom}

In presence of a charge impurity with potential  $V_0 \delta^{(3)}({\bf r})$ and a magnetic impurity with potential $V_1 {\bf S}({\bf r})\cdot\hat{\bf J} \delta^{(3)}({\bf r})$, a Luttinger semimetal is perturbed by the Hamiltonian~[\cite{Luttinger}]
\begin{align}\label{eq:hext}
    \hat{H}_{1}({\bf r}) =  \bigg( V_0  \hat{\mathbbm{1}} + V_1~{\bf S} \cdot \hat{\bf J} \bigg) \delta^{(3)}({\bf r}),
\end{align}
where ${\bf S} = \{ S_x, S_y, S_z \}$. We neglect the anisotropic contributions, such as ${\bf S} \cdot \hat{\bf J}^3$ [\cite{Luttinger}], which is a reasonnable approximation for most Luttinger semimetals with the exception of Pr$_2$Ir$_2$O$_7$~[\cite{Luttinger, taekoo, zhang}]. 

In linear perturbation theory, the carrier density  $J_0({\bf r})  = \langle \hat{n}({\bf r}) \rangle = \langle \hat{\psi}^{\dagger}_{\bf r}\hat{\psi}_{\bf r} \rangle$ and the $j = 3/2$ pseudo-spin operators $J_i({\bf r}) = \langle \hat{J}_i({\bf r}) \rangle = \langle \hat{\psi}^{\dagger}_{\bf r}\hat{J}_i\hat{\psi}_{\bf r} \rangle$ are
\begin{align}
    J_{\mu}({\bf r}) = \sum_{\nu = 0}^{3} \chi_{\mu\nu}({\bf r}) S_{\nu},
\end{align}
where $\mu \in \{0,1,2,3\}$ and $S_{0} = 1$. The generalized susceptibility at a temperature $T$ is a sum over Matsubara frequencies
\begin{align}\label{eq:xir}
    \chi_{\mu\nu}({\bf r},T) = - T \sum_{\omega_n} {\rm Tr}\bigg[ \hat{G}( {\bf r},i \omega_n) \hat{J}_{\mu} \hat{G}( -{\bf r},i \omega_n) \hat{J}_{\nu} \bigg].
\end{align}
Here, for sake of clarity we introduce $\hat{J}_0 \equiv \hat{\mathbbm{1}}$. Note that in the present work we only consider the static regime, \emph{i.e.} Eq. \eqref{eq:xir} does not depend on frequency, which is appropriate in the study of impurities. The dynamic charge polarizability at T = 0 was derived in [\cite{ourwork,mauri}]. In a previous work [\cite{ourwork}] we also show that the Hamiltonian \eqref{eq:h0} has no charge-spin coupling, that is $\chi_{i0} = \chi_{0i} = 0$ for $i \in \{1,2,3\}$. In Weyl semimetals, this coupling between charge and spin degrees of freedom allows for spin polarization of charge fluctuations [\cite{chargespin}]. The generalised susceptibility in the unperturbed, homogeneous gas only depends on $({\bf r} - {\bf r}')$ and $(t - t')$. In Eq.~\eqref{eq:xir} and what follows we express $\chi_{00}$ in units of $V_0 E_F N_0^2$ and $\chi_{ij}$ in units of $V_1 E_F N_0^2$, where $N_0 = 1/(4\pi^2)$ is the density of states per spin of an electron gas.

\begin{figure*}[tb]
    \centering
    \includegraphics[width = \textwidth]{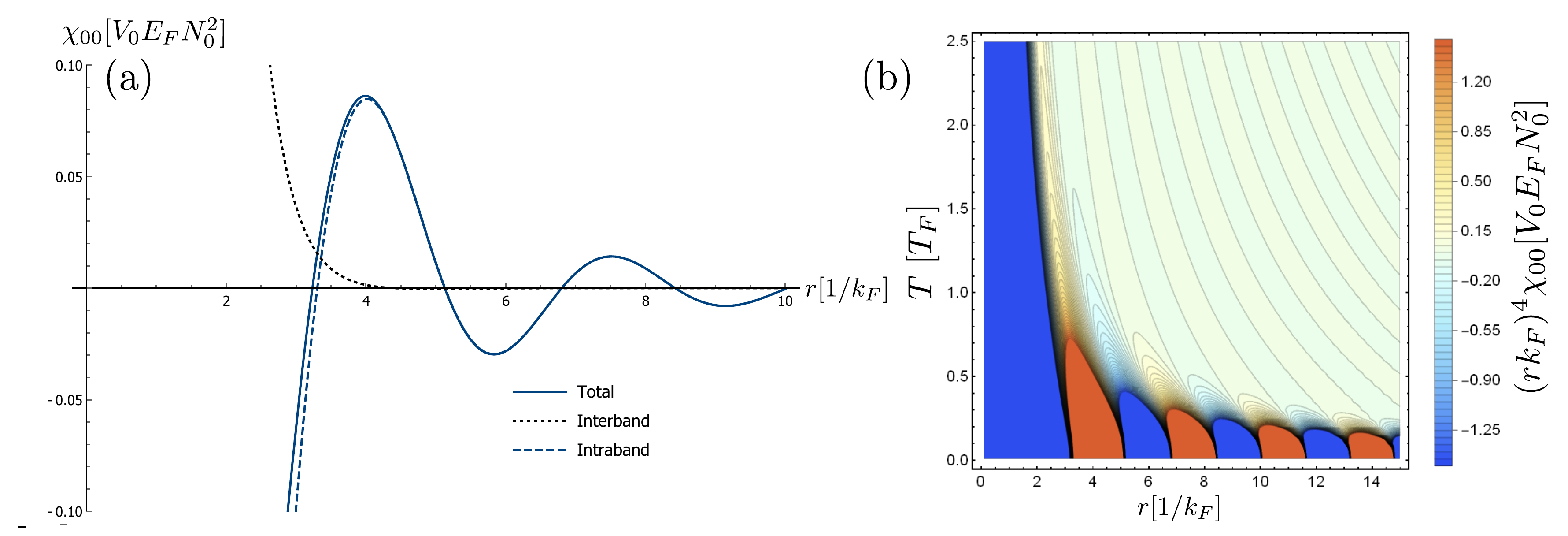}
    \caption{Friedel oscillations in response to an impurity that increases the potential at ${\bf r} = 0$ in a Luttinger semimetal with hole-like doping at (a) $T = 0$ with both intraband and interband contributions (b) $T \neq 0$, showing the damping as the temperature increases. The intraband contribution dominates at both short and long range and the interband coupling is only apparent at short range.}
    \label{fig:friedelcomp}
\end{figure*}

The real space representation of the Green's function in Eq.~\eqref{eq:xir} is derived in Appendix~\ref{app:green}. We use rotation symmetry to absorb the angular dependence in a unitary transformation $\hat{U}_{\theta\phi}$ on the spinor subspace, \emph{i.e.} $\hat{G}({\bf r}, i\omega_n) = \hat{U}_{\theta,\phi}\hat{G}(r {\bf e}_z ,i\omega_n)\hat{U}^{\dagger}_{\theta,\phi}$ and
\begin{widetext}
\begin{align}
    \label{eq:gr}
    \hat{G}(r {\bf e_z}, &i\omega_n)/N_0 = \bigg[ (i\omega_n - {\rm sgn}(m)) I_0(r, i\omega_n) + \frac54 \bigg( \frac{2}{r} { \partial_r I_{0}(r, i\omega_n)} + { \partial_r^2 I_0(r, i\omega_n)} \bigg) \bigg]  \hat{\mathbbm{1}} - \partial_r I_{0}(r, i\omega_n) (\hat{J}^2_x + \hat{J}^2_y) - { \partial_r^2 I_0(r, i\omega_n)} \hat{J}^2_z,
\end{align}
\end{widetext}
where the radial dependence of the Green's function is described by $I_{0}(r, i\omega_n) = - \frac{1}{r}(A_2(r, i\omega_n) + B_2(r, i\omega_n))$ and its derivatives.

We introduce the functions $A_{p}(r, i\omega_n)$ and $B_{p}(r, i\omega_n)$ which are related to the contribution from the valence and from the conduction band for a given sign of $m$ as schematised in Fig. \ref{fig:spctr} :
\begin{align}
    A_{p}(r, i\omega_n) &= \frac{\pi (-i \, {\rm sgn}(\omega_n))^p}{2 (-{\rm sgn}(m)+i\omega_n)^{p/2}} e^{{\rm sgn}(\omega_n) i r \sqrt{-{\rm sgn}(m) + i\omega_n}},\\
    B_{p}(r, i\omega_n) &= \frac{\pi (-i \, {\rm sgn}(\omega_n))^p}{2({\rm sgn}(m)-i\omega_n)^{p/2}} e^{-{\rm sgn}(\omega_n) i r \sqrt{{\rm sgn}(m) - i\omega_n}}.
\end{align}

Thus, the generalized susceptibility \eqref{eq:xir} is a sum over all excitations within or between the two bands, which are decomposed over the functions ${\rm AA}_{k+p}(r, T) = T \sum_{\omega_n > 0} A_{k}A_{p}$, ${\rm BB}_{k+p}(r, T) = T \sum_{\omega_n > 0} B_{k}B_{p}$ and ${\rm AB}_{k+p}(r, T) = i^k T \sum_{\omega_n > 0} A_{k}B_{p}$ : the expressions with the sum on negative frequencies are simply the complex conjugate of these ones. 
We numerically perform the summation on Matsubara's frequencies at $T \neq 0$. At $T = 0$ the sum becomes an integral and we compute explicitly these expressions in Appendix~\ref{app:aabbab}. As expected we find that at zero temperature the intra-valence band terms vanish \emph{i.e.} there are no contributions from intra-valence band scattering since the band is empty. In what follows we separately discuss the charge response, related to Friedel oscillations, and the magnetic response, related to the RKKY coupling. From the last we compute the temperature dependence of the Pauli susceptibility.

In the following we express the susceptibilities in terms of the combinations ${\rm AA}_{p}(r, T)$, ${\rm BB}_{p}(r, T)$ and ${\rm AB}_{p}(r, T)$ for a general value of the mass $m$. In Appendix~\ref{app:aabbab} we explicitly compute the expression of these combinations at $T = 0$ in the case of a positive mass and deduce the corresponding expressions for a negative one.
~~

\subsection{Friedel oscillations}

The charge response of a spin-degenerate normal electron gas at $T = 0$ in units of $V_0 E_F N_0^2$ is [\cite{white}]
\begin{align}\label{eq:normq}
    &\chi^{\rm 3DEG}_{00}(r, T = 0) = \frac{2\pi(\sin(2r) - 2 r \cos(2r))}{ r^4}
\end{align}
In a Luttinger semimetal, the generalized susceptibility also includes interband transitions due to the strong spin-orbit coupling. We compute the trace in Eq.~\eqref{eq:xir} using the Green's function \eqref{eq:gr} and we find 
\begin{widetext}
\begin{align}\label{eq:xi0r}
    \chi_{00}(r, T) = &-\frac{4}{r^6}\bigg( 9( {\rm BB}_4 + {\rm AA}_4 ) + 18 r ( {\rm BB}_3 - {\rm AA}_3 ) + 15 r^2 ({\rm BB}_2 + {\rm AA}_2 ) + 6r^3 ( {\rm BB}_1 - {\rm AA}_1) + 2 r^4 ({\rm BB}_0 + {\rm AA}_0)\nonumber\\
    &~~~~~~~~~~~~  - 18 {\rm AB}_4 - 18 (1 - i ) r {\rm AB}_3 + 18 r^2 i {\rm AB}_2 + 6r^3 (1+i) {\rm AB}_1 \bigg) + {\rm c.c.},
\end{align}
\end{widetext}
where the functions on the right-hand side are evaluated for $(r, T)$ and ${\rm c.c.}$ stands for complex conjugate. We note here that the charge response is symmetric with respect to the sign of the mass.

At zero temperature $T = 0$, the charge susceptibility behaves as in Fig.~\ref{fig:friedelcomp}(a). In this figure we decompose the charge susceptiblity in terms of intraband contributions, given by $AA_k$ or $BB_k$ terms, and interband contributions, given by $AB_k$ terms. The intraband contribution shows an oscillating behaviour, which is expected from the sharpness of the Fermi surface that forbids excitations with wavevector $q > 2 k_F$, in the static regime. Oscillations in the interband contribution decay exponentially with distance since the only interband contributions are virtual excitations in the static regime. At long range, $r\gg 1$, the response is dominated by the intraband contribution and is about half of that in a normal electron gas given in Eq.~\eqref{eq:normq}. The consequences of spin-orbit coupling are more apparent at short-range, $r \ll 1$, where the response is approximately
\begin{align}
    \chi_{00}(r \ll 1, T = 0) \approx -\frac{3\pi(\pi-2)}{ r^4}.
    \label{eq:xi0000}
\end{align}
This is in strong contrast with a normal electron gas where $\chi^{\rm 3DEG}_{00}(r \ll 1, T = 0,) \approx 16 \pi/3r$.
This is the result of a strong intraband coupling at short range as we can see from the intraband and interband contributions of this asymptotic behavior. We find that the intraband contibution is $\chi_{00}^{\rm(intra)} \approx \frac{-3 \pi^2}{r^4} + \frac{8 \pi}{5 r}$ and the interband one is $\chi_{00}^{\rm(inter)} \approx \frac{6 \pi}{r^4} - \frac{8 \pi}{5 r}$ and we notice that the intraband contribution dominates at small $r$. Also, the $1/r$ behavior associated with the 3DEG appears in both contributions, but totally cancel each other. The power law in Eq.~\eqref{eq:xi0000} can be related to the linear component of charge polarisability in momentum space [\cite{ourwork}]. This is of similar origin to the $1/r^5$ divergence observed in Dirac semimetals and the difference in power laws is a consequence of the difference in band dispersions.

At $T \neq 0$ we observe a change in the decaying behavior, from power-law to exponential decay, as can be seen in Fig.~\ref{fig:friedelcomp}(b). There, we also observe the expected damping of Friedel oscillations with a change in their periodicity.

\begin{figure*}[tb]
    \centering
    \includegraphics[width = \textwidth]{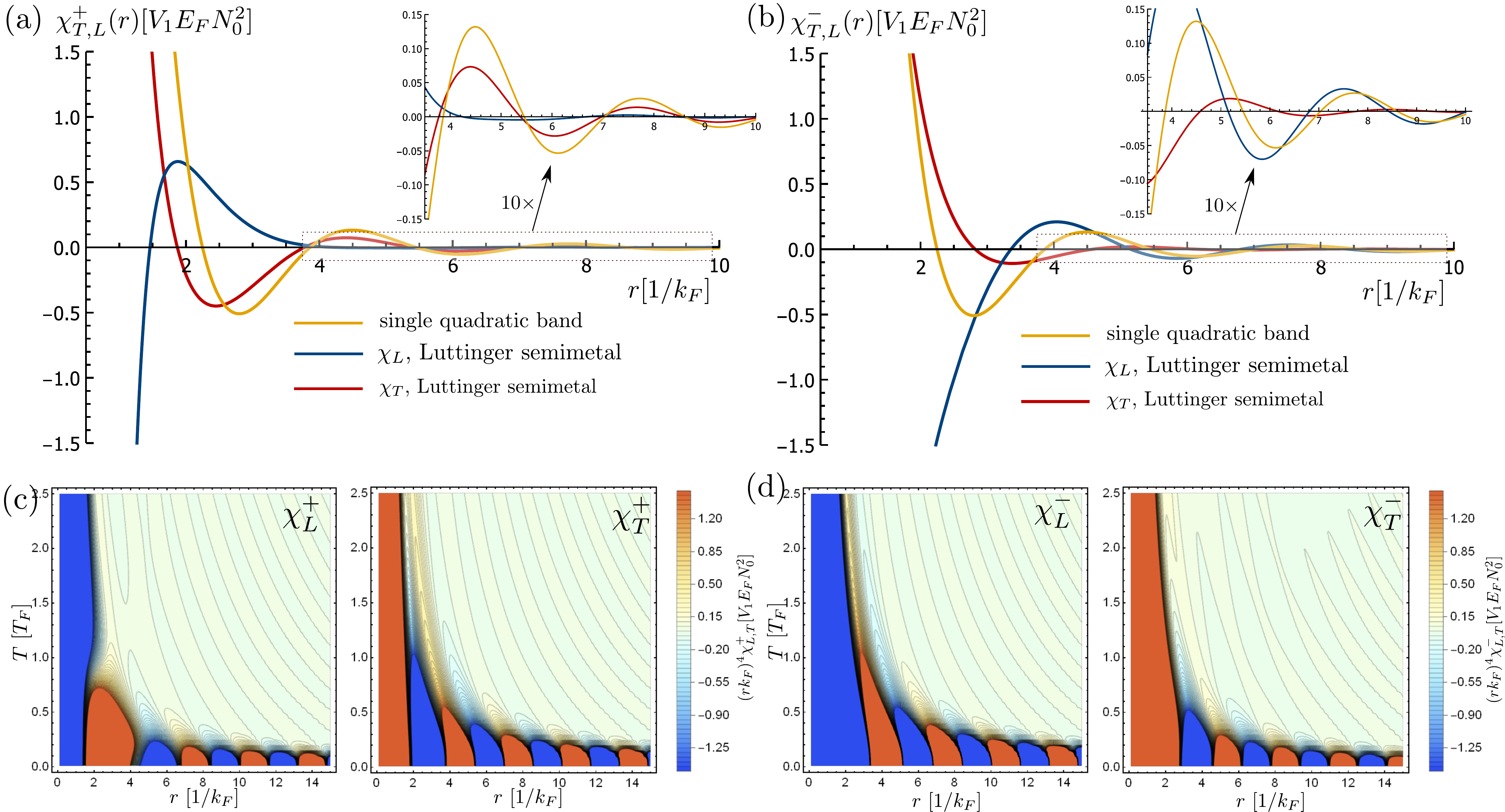}
    \caption{Amplitude of the transverse, $\chi_T$, and longitudinal, $\chi_L$, RKKY coupling between two magnetic impurities for a single quadratic band ($\chi_L = \chi_T$, with $m < 0$) and a Luttinger semimetal (with hole-like doping). (a,b) At $T = 0$ with a $10\times$ zoomed inset, with (a) $m > 0$ and (b) $m < 0$. (c,d) At finite temperature with (c) $m > 0$ and (d) $m < 0$. On the contrary to a normal electron gas, a Luttinger semimetal has opposite transverse and longitudinal couplings at short distances. This behaviour is preserved at larger temperatures where we also observe a change in the periodicity of the oscillations.}
    \label{fig:rkkycomp}
\end{figure*}

\subsection{RKKY coupling}

In absence of spin-orbit coupling, the magnetization profile of a normal electron gas in response to a magnetic impurity is isotropic (\emph{i.e.} $\chi_{ij} = \chi_s^{\rm 3DEG} \delta_{ij}$) just like the charge response in Eq.~\eqref{eq:normq} [\cite{white, mahan}]. On the contrary, in a Luttinger semimetal, the spin-response $\hat{\chi}_S$ is anisotropic. The angular dependence of $\hat{\chi}_S$ is absorbed in the rotation matrix $\hat{R}_{\theta\phi}$, where $\theta,\phi$ are the angular coordinates of the separation ${\bf r}$ to the impurity,
\begin{align} \label{eq:chiS}
    \hat{\chi}_S({\bf r},T) = \hat{R}_{\theta,\phi} 
    \left(
        \begin{array}{ccc}
            \chi_{T}(r, T) & 0 & 0\\
            0 & \chi_T(r, T) & 0 \\
            0 & 0 & \chi_L(r, T)
        \end{array}
    \right)
    \hat{R}_{\theta, \phi}^T.
\end{align}
The diagonal components $\chi_{T}(r, T)$ and $\chi_{L}(r, T)$ are respectively the transverse and longitudinal spin response:
\begin{widetext}
    \begin{align}
    \label{eq:chist}
        &\chi_{T}(r, T) = \frac{1}{r^6}\bigg( 9( {\rm BB}_4 + {\rm AA}_4 ) + 18 r ( {\rm BB}_3 - {\rm AA}_3 ) + 3 r^2 ( {\rm BB}_2 + 9 {\rm AA}_2 ) - 6 r^3 ( {\rm BB}_1 + 3 {\rm AA}_1) - 8 r^4 {\rm BB}_0\\
        &~~~~~~~~~~~~~~~  - 18 {\rm AB}_4 - 18 (1 - i) r {\rm AB}_3 + 6(4+3i) r^2 {\rm AB}_2  + 6 ( 3 - i) r^3 {\rm AB}_1 + 12 r^4 {\rm AB}_0 \bigg) + {\rm c.c.},\nonumber\\
        &\chi_{L}(r, T) = -\frac{1}{r^6}\bigg( 45( {\rm BB}_4 + {\rm AA}_4 ) + 90 r ( {\rm BB}_3 - {\rm AA}_3 ) + r^2 (51 {\rm BB}_2 + 99 {\rm AA}_2 ) + 6 r^3 ( {\rm BB}_1 - 9 {\rm AA}_1) + 2 r^4 ( {\rm BB}_0 + 9 {\rm AA}_0)\nonumber\\
    \label{eq:chisl}
        &~~~~~~~~~~~~~~~  - 90 {\rm AB}_4 - 90 (1 - i) r {\rm AB}_3 + 6(8+15i) r^2 {\rm AB}_2  + 6 ( 9 + i) r^3 {\rm AB}_1 \bigg) + {\rm c.c.},
    \end{align}
\end{widetext}
where the functions on the right-hand side are evaluated for $(r, T)$ and ${\rm c.c.}$ stands for complex conjugate. The functions ${\rm AA}_p$, ${\rm BB}_p$ and ${\rm AB}_p$ are the same as the ones involved in computing the charge response in Eq.~\eqref{eq:xi0r} and their expressions at $T = 0$ are reported in Appendix~\ref{app:aabbab}.

We find that the spin-response depends on the sign of the mass $m$ in Eq.~\eqref{eq:h0}. We remind that this parameter does not affect the band dispersion but only the chirality of the states at the Fermi surface, which we illustrate in Figs.~\ref{fig:spctr}(a,b). We thus distinguish the situations where $m$ is positive and negative with a superscript $\pm$ on the spin-response. In Fig.~\ref{fig:rkkycomp}(a,b) we plot the transverse and longitudinal spin susceptibility at zero temperature. 

The intraband scattering dominates the long range behavior, $r \gg 1$, and the spin response follows the same power law as in a normal electron gas. However the spin-response stays anisotropic even at long range with, for each sign of $m$,
\begin{align}
    &\left\{
    \begin{array}{l}
        \chi_{T}^+(r \gg 1, T = 0) \approx \chi^{\rm 3DEG}_{00}(r)/2\\
        \chi_{L}^+(r \gg 1, T = 0) \approx \chi^{\rm 3DEG}_{00}(r)/8
    \end{array}
    \right.,\\
    &\left\{
    \begin{array}{l}
        \chi_{T}^-(r \gg 1, T = 0) \approx -\frac{9 \pi}{2 r^4} \sin(2 r)\\
        \chi_{L}^-(r \gg 1, T = 0) \approx 9\chi^{\rm 3DEG}_{00}(r)/8
    \end{array}
    \right. .
    \label{eq:rkkylarger}
\end{align}
We notice that in the long-range behaviour, the longitudinal response has an amplitude $\lambda^2/2$, with $\lambda$ the helicity crossing the Fermi surface (see Fig.~\ref{fig:spctr}), compared to the normal electron gas. This does not happen for the transverse response, where we even observe that $\chi_T^-$ decreases in $1/r^4$ instead of $1/r^3$ for the normal electron gas.

Close to the magnetic impurity, the short range spin-spin response is anisotropic with opposite transverse and longitudinal contributions
\begin{equation}\label{eq:rkkysmallr}
    \begin{split}
    \chi_T^\pm(r \ll 1, T=0) &\approx 
    \frac{3\pi(2+\pi)}{4 r^4} \mp \frac{2 \pi}{3 r} ,\\
    \chi_L^\pm(r \ll 1, T=0) &\approx - \frac{15 \pi(\pi-2)}{4 r^4} \pm \frac{4\pi}{3 r} .
    \end{split}
\end{equation}
This, again, is related to the strong spin-orbit coupling and opens the possibility to observe both ferromagnetic and antiferromagnetic coupling between magnetic impurities, from the RKKY coupling. 

We observe that for a positive mass, $m > 0$, the contribution to the transverse spin-response $\chi_T^+$ from intraband and interband coupling are respectively $\frac{15 \pi^2}{4 r^4} - \frac{6 \pi}{r^3} + \frac{8 \pi}{5 r}$ and $\frac{3 \pi (1-2\pi)}{2 r^4} + \frac{6 \pi}{r^3} - \frac{34 \pi}{15 r}$, and the corresponding behaviors in the longitudinal component $\chi_L^+$ are $-\frac{39 \pi^2}{4 r^4} + \frac{12 \pi}{r^3} + \frac{2 \pi}{r}$ and $\frac{3 \pi(5 + 4\pi)}{2 r^4} - \frac{12 \pi}{r^3} - \frac{2 \pi}{3 r}$. The intraband contribution then always dominates close to the magnetic impurity. On the contrary, for a negative mass, $m < 0$, the intraband and interband contributions to $\chi_T^-$ are respectively $\frac{- 9 \pi^2}{4 r^4} + \frac{6 \pi}{r^3} + \frac{8 \pi}{5r}$ and $\frac{3\pi(1+2\pi)}{2 r^4} - \frac{6 \pi}{r^3} - \frac{14 \pi}{15 r}$ and for $\chi_L^-$ they are respectively $\frac{9 \pi^2}{4 r^4} - \frac{12 \pi}{r^3} + \frac{2 \pi}{r }$ and $\frac{3 \pi (5 - 4 \pi)}{2 r^4} + \frac{12\pi}{r^3} - \frac{10 \pi}{3 r}$. Then, close to the impurity, the two spin-responses are instead dominated by the interband contribution. This indicates that the magnetic coupling at small separation is dominated by excitations involving the bands with the smallest helicities, $\lambda = \pm 1/2$.

Similar to our observation for Friedel oscillations, this peculiar behaviour at proximity to the magnetic impurity persists at larger temperatures as shown in Figs.~\ref{fig:rkkycomp}(a,b). As the temperature increases, we observe a decrease in the periodicity of the RKKY oscillations and a decay in their amplitude. The long range behavior of the spin oscillations shows an exponential decay at finite temperature. In order to complete this discussion, we now obtain the effective RKKY Hamiltonian between two impurities and also compute the Pauli susceptibility in a Luttinger semimetal.

\subsection{Effective RKKY Hamiltonian}\label{subsect:Hamil}

The interaction between two magnetic impurities ${\bf S}_1$ and ${\bf S}_2$ localized at respectively ${\bf r}_1$ and ${\bf r}_2$ is described by the coupling Hamiltonian $\hat{H}_{12}$:
\begin{align}
    \hat{H}_{12} = V_1 \, {\bf S}_1^T \hat{\chi}_S({\bf r}, T) {\bf S}_2,
\end{align}
where ${\bf r}={\bf r}_2-{\bf r}_1$, and $\hat{\chi}_S({\bf r}, T)$ is given in Eq.~\eqref{eq:chiS}. This can be rewritten in a more explicit way:
\begin{align}\label{eq:Heffective}
    \hat{H}_{12}/ V_1 = \chi_T {\bf S}_1\cdot {\bf S}_2 + (\chi_{L} - \chi_T) ({\bf S}_1\cdot {\bf e}_r) ({\bf S}_2 \cdot {\bf e}_r),
\end{align} 
where ${\bf e}_r$ is the unit vector that separates the two impurities and $\chi_{L,T}$ are evaluated for $r = |{\bf r}_2 - {\bf r}_1|$, the distance between the magnetic impurities. The first and second terms in Eq.~\eqref{eq:Heffective} are respectively the Heisenberg and Ising contributions. There is no Dzyaloshinskii-Moriya interaction which would be a consequence of asymmetric spin-orbit coupling~[\cite{imamura04,wang17}] and which is absent in Eq.~\eqref{eq:h0}.

This effective spin-spin coupling Hamiltonian differs from that in a normal electron gas where it is Heisenberg-like, \emph{i.e.} only the first contribution in Eq.~\eqref{eq:Heffective} is present. Here, in the case of Luttinger semimetals, we obtain an additional coupling between the spin components parallel to their separation and we evaluate the amplitude of this term for various Luttinger semimetals in the next section. We also note that in Ref.~[\cite{flint}], the coupling between the spin chiralities of a Luttinger semimetal is described by a Heisenberg Hamiltonian.

\subsection{Spin susceptibility}\label{subsect:Pauli}

In presence of a uniform magnetic field, ${\bf B}$, the Zeeman coupling will be~[\cite{Luttinger}]
\begin{align}
    \hat{H}_2 = - g \mu_B \hat{\bf J}\cdot {\bf B},
\end{align}
where we introduce the $g$-factor and the Bohr magneton $\mu_B = e/(2m_e c)$. The spin magnetization ${\bf M} \equiv - \nabla_{\bf B} \langle \hat{H}_2 \rangle = g \mu_B \langle \hat{\bf J} \rangle = \hat{\chi}_P {\bf B}$ defines the Pauli susceptibility $\hat{\chi}_P$ [\cite{giuliani}]
\begin{equation}\label{eq:chip}
\begin{split}
    \hat{\chi}_P(T) &\equiv \lim_{k\rightarrow 0} \frac{\hat{\chi}_S({\bf k}, T)}{4\pi^2}  = \frac{1}{4\pi^2}\lim_{k\rightarrow 0} \int d^3{\bf r} ~\hat{\chi}_S({\bf r}, T) e^{-i{\bf k}\cdot{\bf r}} \\
    &= \frac{1}{3 \pi}\int_{0}^{\infty}dr~r^2 \big(2\chi_T(r, T) + \chi_L(r, T)\big) \hat{\mathbbm{1}},
\end{split}
\end{equation}
that we write in units of $(g\mu_B)^2 N_0$. Here, we perform the angular integral in position-space and take the limit $k \rightarrow 0$.

\begin{figure}[tb]
    \centering
    \includegraphics[width = \columnwidth]{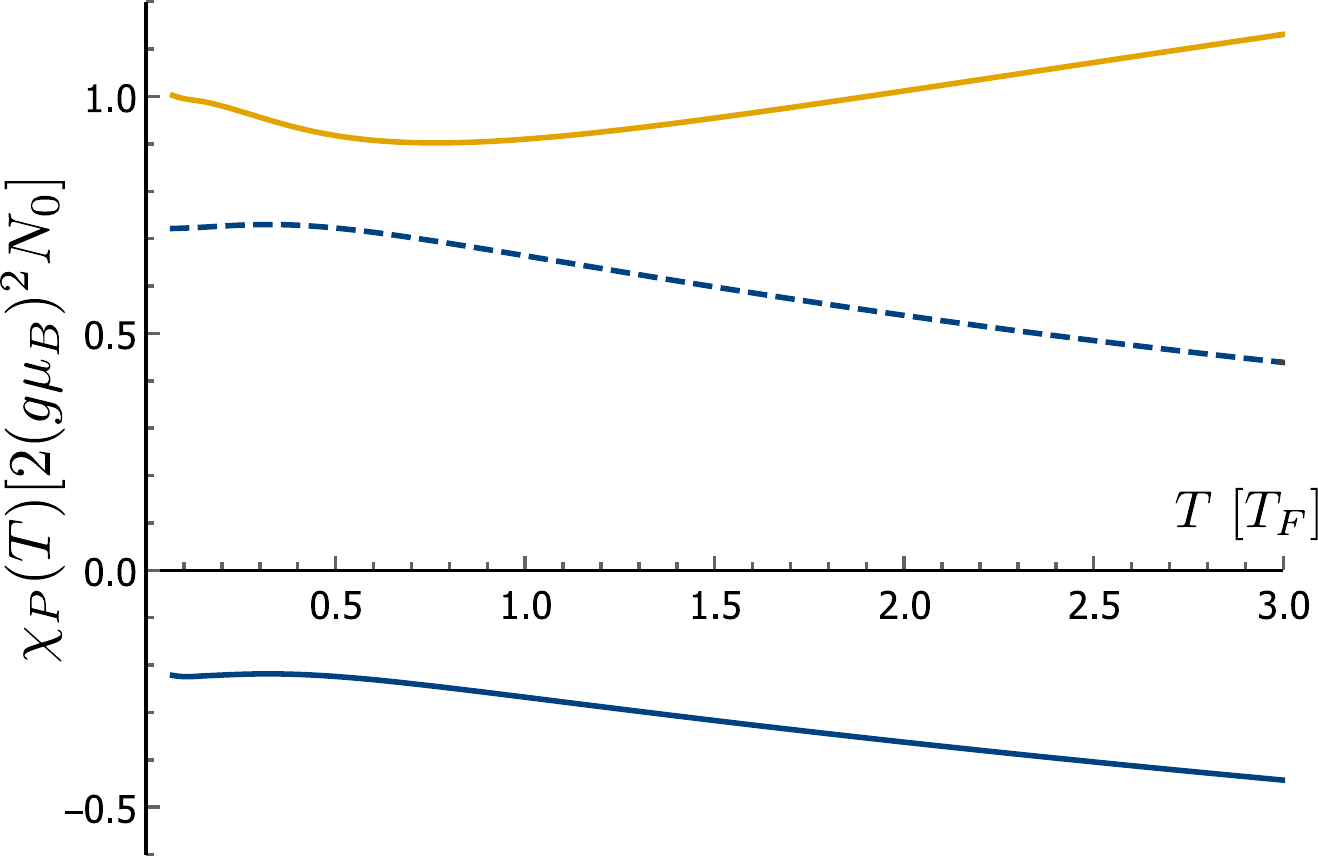}
    \caption{Magnetic susceptibilities as a function of temperature. The Pauli susceptibilities of a normal electron gas (in orange) is paramagnetic while it is diamagnetic in a Luttinger semimetal (plain line, in blue). 
    Also shown is the Landau susceptibility in Luttinger semimetals [\cite{leahy}] that we draw for a cutoff energy $E_0/E_F = 10$ (dashed line).
    }
    \label{fig:chit}
\end{figure}

We compute this integral numerically after subtracting the $1/r^4$ asymptotic contribution in Eq.~\eqref{eq:rkkysmallr} from $\chi_T$ and $\chi_L$. These terms do not contribute to the Pauli susceptibility [\cite{divnote}] and by substracting them we avoid numerical instabilities. We obtain the behavior reported in Fig.~\ref{fig:chit} and compare it to the Pauli susceptibility of a normal spin-degenerate electron gas. The response is diamagnetic instead of being paramagnetic, which is analogous the unusual paramagnetic Landau susceptibility in Luttinger semimetals [\cite{leahy}] that we reproduce in Fig.~\ref{fig:chit} for a cutoff energy $E_0/E_F = 10$. This diamagnetic Pauli susceptibility is a consequence of interband transitions and we find that the contribution of intraband and interband excitations to the susceptibility $\chi_P = \chi_P^{\rm (intra)} + \chi_P^{\rm (inter)}$ is, at $T = 0$, $\chi_P^{\rm (intra)}  = 5\chi_{P}^{\rm 3DEG}/4$ and $\chi_P^{\rm (inter)} = -3\chi_P^{\rm 3DEG}/2$ where $\chi_{P}^{\rm 3DEG}$ is the spin susceptibility of the normal electron gas. We also notice that the Pauli susceptibility is independent of the sign of the mass. This behavior is also drastically different from that in Weyl semimetals where the Pauli susceptibility cancels because a magnetic field only splits the cones in momentum space [\cite{chidirac}] and where the Landau susceptibility is diamagnetic [\cite{chidirac,leahy}].

\section{Discussion}
\label{sec:discussion}

The carrier density of Luttinger semimetals, such as YPtBi [\cite{yptbi1,yptbi2,yptbi3,yptbi4}] and Pr$_2$Ir$_2$O$_7$ [\cite{priro1}] is of the order of $n \approx (0.1 - 1)\times 10^{19}$ cm$^{-3}$, that is $k_F \approx 0.3-0.7~{\rm nm}^{-1}$ which is comparable to the lattice constant, $a \approx 0.5-1~{\rm nm}$~[\cite{Nakajima}]. Thus $a k_F \approx 0.1-1$ which is well within the region with unusual charge and spin responses due to spin-orbit coupling. With impurities separated by the lattice constant size, one can expect to see the stronger charge coupling near a charge impurity and opposite signs for the transverse and longitudinal spin-spin interaction. Note that YPtBi lacks a center of inversion and allows for additional asymmetric spin-orbit interactions in Eq.~\eqref{eq:h0}, that split the bandstructure. In the present discussion we neglect this effect which is not always present [\cite{yptbi4}] and leads to a Fermi surface splitting $\Delta k$ of at most $8\%$ of $k_F$ in this material [\cite{yptbi1}].
More specifically, in the case of Pr$_2$Ir$_2$O$_7$ the magnetic coupling occurs between the $4f$ orbitals of the Pr$^{3+}$ ions described by $J = 4$ magnetic non-Kramers doublets with $J_z = \pm 4$ [\cite{machida}] and separated by a distance $a k_F \approx 1.7-2$ [\cite{takatsu,kondo,cheng}]. Then, the ratio between the two contributions in the effective Hamiltonian \eqref{eq:Heffective}, given by $ \frac{\chi_L - \chi_T}{\chi_T} $ is in the region $[-2.5, -3]$ if one takes $m < 0$. In this interval of $a k_F$, one coupling is ferromagnetic while the other is antiferromagnetic, depending on the sign of $V_1$ as can be seen in Fig.~\ref{fig:rkkycomp}(b).

However this ratio strongly fluctuates and can reach large values if one takes $m > 0$ since $\chi_T$ is close to zero according to our calculations (see  Fig. \ref{fig:rkkycomp}(a)). In this case $\chi_T$ can be positive or negative in this interval and $\chi_L - \chi_T > 0$. A similar analysis for the family of bismuth half-Heusler MXBi (M = Y, Lu) (X = Pd, Pt) where the ions are separated by $a k_F \approx 0.1-0.5$ [\cite{Nakajima, wang, pavlosiuk}], gives ratios independent on the sign of the mass and of about $-2$, again with one coupling being ferromagnetic and the other antiferromagnetic.
In a recent work we have pointed the importance of these Friedel oscillations in the superconductivity from the Coulomb repulsion in Luttinger semimetals [\cite{scus,tchoumakov2019interplay}]. The anisotropic magnetic coupling may, as well, be responsible for exotic magnetic phases [\cite{spinice,krempa}].
However, these short-range effects may be strongly renormalized by interactions. Here, we neglect screening and vertex corrections which are the strongest at short-range and may push these effects to even shorter-ranges, as in the normal electron gas [\cite{friedelrpa}].

The consequences of spin-orbit coupling on Friedel and RKKY oscillations in Luttinger semimetals have some similarities with isotropic Dirac semimetals. Dirac semimetals also have a strong spin-orbit coupling but with a linear band dispersion that leads to a $1/r^3$ decay of the oscillations [\cite{friedeldirac,rkkydirac2,rkkydirac4, hosseini}]. These semimetals also show a spin-spin coupling with Heiseinberg and Ising contributions as in \eqref{eq:Heffective} and similar structures [\cite{kurilovich2016,kurilovich2017,verma2019}] and that can be made more anisotropic if the Weyl cones are split [\cite{rkkydirac2,rkkydirac4, hosseini}]. This anisotropy also survives at long-range~\eqref{eq:rkkylarger} and at higher temperatures,~Fig.~\ref{fig:rkkycomp} (c,d). The Weyl cones are characterized by an helicity operator, related to the cones chirality, but the RKKY coupling is independent on this degree freedom. This is in stark contrast to the behavior observed in our calculations for Luttinger semimetals, where the helicity at the Fermi surface matters for the profile of magnetic coupling.

It is also interesting to take the limit $\mu = 0$ at $T = 0$, in which case the carrier density vanishes and the model does not possess any energy scale. In Dirac and Weyl semimetals, this result in a nonoscillatory $r^{-5}$ decaying RKKY behavior [\cite{rkkydirac2,rkkydirac4, hosseini}]. In the case of Luttinger semimetal, the Friedel and RKKY responses show a nonoscillatory $r^{-4}$ decaying behavior and we also note that the response is exactly the $r^{-4}$ terms from the expressions in Eq.~\eqref{eq:xi0000} for Friedel oscillations and Eq.~\eqref{eq:rkkysmallr} for the RKKY responses. This difference in power law between Dirac and Luttinger semimetals could prove useful to experimentally probe the strong orbit-coupling since the response is more long-ranged in Luttinger semimetals. In the previous subsection \ref{subsect:Pauli} we also find a diamagnetic Pauli susceptibility, whereas in Weyl semimetals where the Pauli susceptibility cancels [\cite{chidirac}]. Since the two band structures can be related by applying strain or a magnetic field [\cite{luttodirac,Szabo,ghorashi1,ghorashi2}], it is interesting to see that our results should still hold for energies larger than the energy scale of the Weyl cones. GdPtBi is an example of such a material, with induced Weyl point from the rare-earth exchange field, but with quadratic band dispersion far from the Fermi surface [\cite{liu,shekhar}].

~~\\
\section{Conclusion}
In this work we compute the response of a Luttinger semimetal to a charged and to a magnetic impurity. At large separations, the charge and magnetic oscillations are similar to that of a normal electron gas, up to an anisotropy for the magnetic response. The main difference between the two systems is at short distance, where spin-orbital effects are the most important and result in a $r^{-4}$ divergence. In particular we observe opposite transverse and longitudinal magnetic couplings, even if the model is isotropic. We obtain the RKKY interaction Hamiltonian between two impurities and compute the Pauli susceptibility and find that it is diamagnetic instead of being paramagnetic, which is in line with previous calculations that find a paramagnetic orbital susceptibility in Luttinger semimetals [\cite{leahy}].

This response of Luttinger semimetals to impurities may lead to exotic phase transitions [\cite{nandkishore,mandal}]. The Friedel oscillations contribute to the Kohn-Luttinger mechanism of superconductivity [\cite{kohnluttinger}] which we have recently studied for Luttinger semimetals [\cite{scus}]. The peculiar RKKY coupling could be at the origin of exotic magnetic phases. In this work we focused on the bulk response of Luttinger semimetals which may have a different behavior at their surface and is more relevant in scanning tunneling microscopy (STM). Indeed, it was recently discussed that these materials may have surface states [\cite{surfaceLuttinger}] with various band dispersions and that could be responsible for a different surface response than described in the present manuscript.

While finishing the present work, we became aware of a similar one that has been published [\cite{mohammadi}]. In this work the authors explore the RKKY interaction in 2D Luttinger's systems with anisotropic electron-hole dispersion, in contrast to the present work where we focus on 3D system and where we observe anisotropy in the RKKY response even with electron-hole symmetry.

\begin{acknowledgments}
This project is funded by a grant from Fondation Courtois, a Discovery Grant and Canada Graduate Scholarships – Master’s Program from NSERC, a Canada Research Chair, and a ``\'Etablissement de nouveaux chercheurs et de nouvelles chercheuses universitaires'' grant from FRQNT. 
\end{acknowledgments} 

\bibliographystyle{apsrev4-1}
\bibliography{bibliography}

\newpage
\appendix 

\begin{widetext} 

\section{Fourier transform of the thermal Green's function}\label{app:green}

In this appendix, we compute the Fourier transform of the thermal Green's function \eqref{eq:green} for general $m$ 
\begin{align} \label{eq:A1}
    \hat{G}({\bf r}, i\omega_n) \equiv  \frac{N_0}{2 \pi} \int d^3{\bf k}~\hat{G}({\bf k}, i\omega_n) e^{i{\bf k}\cdot {\bf r}} =  -\frac{N_0}{2 \pi} \int d^3{\bf k}~\frac{ \left( i\omega_n - {\rm sgn}(m) - \frac54 {\bf k}^2 \right) \hat{\mathbbm{1}} + \left({\bf k}\cdot \hat{{\bf J}}\right)^2}{(\xi_{+}({\bf k}) + i\omega_n)(\xi_{-}({\bf k}) + i\omega_n)} e^{i{\bf k}\cdot {\bf r}}.
\end{align}
with $\xi_{\pm}({\bf k}) = \pm k^2 - {\rm sgn}(m)$. We write \eqref{eq:A1} in terms of the auxiliary integral $I_0(r, i \omega_n)$ and its derivatives $I_{ij}({\bf r}, i\omega_n) = \frac{\partial^2 I_0(r, i\omega_n)}{\partial r_i \partial r_j}$ :

\begin{align}
    \hat{G}({\bf r}, i\omega_n) =  N_0 \bigg[ \left( (i\omega_n - {\rm sgn}(m)) I_0(r, i\omega_n) - \frac54 \sum_{i = 1}^{3} I_{ii}( {\bf r}, i\omega_n) \right) \hat{\mathbbm{1}}  + \sum_{i,j=1}^{3} I_{ij}({\bf r}, i\omega_n) \hat{J}_i \hat{J}_j \bigg],
    \label{eq:app:gr}
\end{align}

The integral $I_0(r, i \omega_n)$ corresponds to the $k$-independant part of the numerator in \eqref{eq:A1} and evaluates to :
\begin{align}
    I_0(r, i\omega_n) &= \int_0^{\infty} dk ~ k^2 \int_{-1}^{1}du \frac{e^{i k r u}}{(k^2 - ({\rm sgn}(m) - i\omega_n))(k^2 - (i\omega_n - {\rm sgn}(m)))}\\
    &= \frac{-i}{ r}\int_{-\infty}^{\infty} dk \frac{k e^{i k r}}{(k^2 - ({\rm sgn}(m) - i\omega_n)) (k^2 - (i\omega_n - {\rm sgn}(m)))}\\
    &= \frac{\pi}{2 r (-{\rm sgn}(m)+i\omega_n)} \bigg( e^{{\rm sgn}(\omega_n) i \sqrt{-{\rm sgn}(m) + i\omega_n} r} - e^{-{\rm sgn}(\omega_n) i \sqrt{{\rm sgn}(m) - i\omega_n} r} \bigg).
\end{align}
The integration of the terms of the form $k_i k_j$ in the numerator of the Green's function can be written as the derivatives of this auxiliary integral :
\begin{align}
    I_{ij}( {\bf r}, i\omega_n) &= -\frac{\partial^2 I_0(r, i\omega_n)}{\partial r_i \partial r_j} = \int_0^{\infty} dk~k^2 \int_{-1}^{1}du \frac{k_i k_j e^{i k r u}}{(k^2 - ({\rm sgn}(m) + i\omega_n))(k^2 + ({\rm sgn}(m) + i\omega_n))}\\
    &= - \frac{r_i r_j}{r^2}\bigg( \frac{\partial^2 I_0(r, i\omega_n)}{\partial r^2} - \frac{1}{r}\frac{\partial I_0(r, i\omega_n)}{\partial r} \bigg) - \frac{\delta_{ij}}{r} \frac{\partial I_0(r, i\omega_n)}{\partial r}.
\end{align}We use rotation symmetry to simplify \eqref{eq:A1} and introduce the unitary transformation $\hat{U}_{\theta\phi}$ generated by the pseudo-spin $j = 3/2$ operator $\hat{\bf J}$ to write $\hat{G}({\bf r}, i\omega_n) = \hat{U}_{\theta\phi} \hat{G} (r{\bf e}_z, i\omega_n)\hat{U}^{\dagger}_{\theta\phi}$. The Green's function $\hat{G}(r{\bf e}_z, i\omega_n)$ can then be similarly decomposed into
\begin{align}
    \hat{G}(r{\bf e}_z, i\omega_n) = N_0 \bigg[ \left( (i\omega_n - {\rm sgn}(m)) I_0(r, i\omega_n) - \frac54 \sum_{i = 1}^{3} I_{ii}(r{\bf e}_z, i\omega_n) \right) \hat{\mathbbm{1}}  + \sum_{i,j=1}^{3} I_{ij}(r{\bf e}_z, i\omega_n) \hat{J}_i \hat{J}_j \bigg],
    \label{eq:app:gr}
\end{align} where we write $I_{ij}(r{\bf e}_z, i \omega_n)$ in a matrix form
\begin{align}
    I_{ij}(r{\bf e}_z, i\omega_n) = \left(
        \begin{array}{ccc}
            - \frac{1}{r}\frac{\partial I_0(r, i\omega_n)}{\partial r } & 0 & 0 \\
            0 & - \frac{1}{r} \frac{\partial I_0(r, i\omega_n)}{\partial r } & 0 \\
            0 & 0 & -\frac{\partial^2 I_0(r, i\omega_n)}{\partial r^2}
        \end{array} 
    \right).
\end{align}
We substitute these expressions in \eqref{eq:app:gr} and obtain the real space Green's function reported in main text,  \eqref{eq:gr}.

\section{Explicit contributions to susceptibilities at $T = 0$, $m > 0$}
\label{app:aabbab}

The generalized susceptibility \eqref{eq:xir} depends on multiple contributions from intra and interband scattering. In this section we explicitly denote the combinations of the $m > 0$ case with a plus superscript, and in the following subsection obtain the relationship between these expressions and those corresponding to the case $m < 0$, denoted with a minus superscript. In the limit $T \rightarrow 0$, with $m > 0$, the intra-valence band contributions ${\rm AA}_p^+(r, T = 0)$ in Eqs. (\ref{eq:xi0r},\ref{eq:chist},\ref{eq:chisl}) vanish $\forall ~ p$ and the Matsubara sums
\begin{equation}
    \begin{split}
    {\rm BB}_{k+p}^+ (r, T) &= T \sum_{\omega_n > 0} B_{k}^+(r, i\omega_n)B_{p}^+(r, i\omega_n),\\
    {\rm AB}_{k+p}^+(r, T) &= i^k T \sum_{\omega_n > 0} A_{k}^+(r, i\omega_n)B_{p}^+(r, i\omega_n),
    \end{split}
\end{equation}
can be evaluated analytically with the Euler-MacLaurin summation formula $T \sum_{\omega_n} \approx \frac{1}{2\pi} \int_{-\infty}^\infty d\omega$. These expressions are
\begin{align} 
    &{\rm BB}_{0}^+(r, T = 0) = \frac{1}{2 \pi} \frac{\pi^2}{4} \int_{0}^\infty d\omega ~  e^{ -2i r \sqrt{1 - i\omega}}  = \frac{\pi(-i + 2r)}{16 r^2} e^{ - 2 i r} , \label{eq:B2} \\
    &{\rm BB}_{1}^+(r, T = 0) = \frac{1}{2 \pi} \frac{\pi^2}{4} \int_{0}^\infty d\omega ~ \frac{e^{ -2 i r \sqrt{1 - i\omega}}}{i\sqrt{1 - i\omega}} = -\frac{ i \pi \, e^{-2 i r}}{8 r},  \\
    &{\rm BB}_{2}^+(r, T = 0) = -\frac{1}{2 \pi} \frac{\pi^2}{4} \int_{0}^\infty d\omega ~ \frac{e^{ -2 i r \sqrt{1 - i\omega}}}{1 - i\omega} =\frac{\pi(-\pi + i \, \text{Ei}(-2 i r))}{4},  \\
    &{\rm BB}_{3}^+(r, T = 0) = \frac{1}{2 \pi} \frac{\pi^2}{4} \int_{0}^\infty d\omega ~ \frac{i e^{- 2 i r \sqrt{1 - i\omega}}}{(1 - i\omega)^{3/2}} = -\frac{\pi( e^{-2 i r} - 2 \pi r + 2 i r \, \text{Ei}(-2 i r))}{4},\\
    &{\rm BB}_{4}^+(r, T = 0) =  \frac{1}{2 \pi} \frac{\pi^2}{4} \int_{0}^\infty d\omega ~ \frac{ e^{- 2 i r \sqrt{1 - i\omega}}}{(1 - i\omega)^{2}} = \frac{ \pi(e^{- 2 i r} (i + 2 r) - 4 r^2 (\pi - i \, \text{Ei}(-2 i r)))}{8}, \label{eq:B6}
\end{align}
where $\text{Ei}(r)$ is the exponential integral and, using that,
\begin{align}
    {\rm AB}_{p}^+(r, T) &= T\sum_{\omega_n > 0} \frac{\pi^2}{4 }\frac{(-i)^{p}}{(1-i\omega_n)^{p/2}} e^{- 2 i  \sqrt{1 - i\omega_n} [( 1 - i)r/2]} = {\rm BB}_p^+((1-i)r/2, T),
\end{align}
one can deduce from (\ref{eq:B2}-\ref{eq:B6}) the corresponding expressions of ${\rm AB}_{p}^+(r, T = 0)$.

\subsection{Relationships to $m < 0$, $\forall$ $T$}
\label{app:aabbabm}

One can obtain the following relationships between the expressions corresponding to a positive mass and the ones for a negative mass : 

\begin{align}
    {\rm AA}_p^-(r, T) = (-1)^p ({\rm BB}_p^+(r, T))^*, \\
    {\rm BB}_p^-(r, T) = (-1)^p ({\rm AA}_p^+(r, T))^*, \\
    {\rm AB}_p^-(r, T) = (-i)^p ({\rm AB}_p^+(r, T))^*
\end{align}where the minus/plus superscript refers to the case $m < 0$ and $m > 0$, respectively, and the asterisk denotes the complex conjugate. Then, for $m < 0$, the intra-valence band contributions are associated to ${\rm BB}_p^-$ and do not contribute to Eqs. (\ref{eq:xi0r},\ref{eq:chist},\ref{eq:chisl}) at $T = 0$ since it is empty.

\end{widetext}
\end{document}